\documentclass[aps,prb,twocolumn,superscriptaddress,showpacs]{revtex4-1}

\usepackage{epsfig}
\usepackage{graphicx}
\usepackage{amsfonts}
\usepackage[figuresright]{rotating}
\usepackage{float}
\usepackage{amssymb}
\usepackage{amsmath}
\usepackage{amsthm}
\usepackage{wasysym}
\usepackage{hyperref}

\theoremstyle{plain}

\newtheorem*{theorem*}{Theorem}

\def\avg#1{\langle#1\rangle}

\def\be{\begin{equation}}
\def\ee{\end{equation}}
\def\bea{\begin{eqnarray}}
\def\eea{\end{eqnarray}}

\def\nn{\nonumber}

\begin{document}
\title{Topological nodal Cooper pairing in doped Weyl metals}
\author{Yi Li}
\affiliation{Princeton Center for Theoretical Science, Princeton
University, Princeton, NJ 08544, USA}
\author{F. D. M. Haldane}
\affiliation{Department of Physics, Princeton University, Princeton,
NJ 08544, USA}
\date{November 10, 2015}

\begin{abstract}
We generalize the concept of Berry connection of the single-electron band
structure to the two-particle Cooper pair states between two Fermi
surfaces with opposite Chern numbers.
Because of underlying Fermi surface topology, the pairing Berry phase
acquires non-trivial monopole structure.
Consequently, pairing gap functions have the topologically-protected
nodal structure as vortices in the momentum space with the total vorticity
solely determined by the monopole charge $q_p$.
The pairing nodes behave as the Weyl-Majorana points of the Bogoliubov-de
Gennes pairing Hamiltonian.
Their relation with the connection patterns of the surface modes from the Weyl band
structure and the Majorana surface modes inside the pairing
gap is also discussed.
Under the approximation of spherical Fermi surfaces, the pairing symmetry
are represented by monopole harmonic functions.
The lowest possible pairing channel carries angular momentum
number $j=|q_p|$, and the corresponding gap functions are holomorphic or
anti-holomorphic functions on Fermi surfaces.
\end{abstract}

\pacs{74.20.Rp,73.43.-f,03.65.Vf}
\maketitle


The study of topological states has renewed our understanding of
condensed matter physics.
The discovery of two-dimensional integer quantum Hall states
\cite{klitzing1980,tsui1982} initiated  the exploration of novel
states characterized by band topology rather than symmetry
\cite{thouless1982,haldane1983,avron1983,niu1985,kohmoto1985,avron1990},
with magnetic band structures that possess non-trivial Chern numbers
arising from  broken time-reversal (TR) symmetry.
The study of Berry curvature of Bloch bands in such lattice structures has
led to to many results in anomalous Hall and quantum anomalous Hall
physics \cite{haldane1988,karplus1954,jungwirth2002,onoda2002,
haldane2004,nagaosa2010,yu2010,chang2013}.
The band structure topology has also been generalized to systems of
topological insulators with TR symmetry \cite{kane2005,kane2005a,
bernevig2006a,bernevig2006,fu2007,fu2007a,moore2007,qi2008,roy2009,
hasan2010,qi2011}.
The stable gapless surface modes which appear at
the boundary of gapped topological systems have analogs in
gapless semi-metallic systems, which  can also have   non-trivial band topology.
For example, topological Weyl semi-metals have been proposed and
realized in three-dimensional (3D) systems in the absence of either TR or
inversion symmetry \cite{murakami2007,wan2011,burkov2011,yang2011,meng2012,witczak-Krempa2012,cho2012, hosur2012,fang2012,son2013,
hosur2013,vazifeh2013, nandkishore2014,hosur2014,potter2014,
haldane2014,wei2014, weng2015,burkov2015,xu2015,xu2015a,
lv2015,lv2015a,bednik2015,jian2015,lu2015,borisenko2015,
qi2015,xiong2015a}.
Their band structure is characterized by degenerate Weyl points in
the Brillouin-zone (BZ), which can be  understood as monopole sources
and sinks of Berry-curvature flux in  $\mathbf k$-space.

Topological phenomena are usually understood in terms of  contributions from
all the filled electronic states rather than the states in the vicinity
of Fermi surfaces.
The apparent disagreement with the central tenet of  Fermi-liquid theory that all
conduction processes can be understood at the Fermi level can be
resolved by introducing the Berry phase of quasiparticles
on  the Fermi surface \cite{haldane2004}.
So far, the study of the Fermi surface topology and the associated
Berry phase structure has mainly been discussed at the single-particle
level  \cite{jungwirth2002,onoda2002,nagaosa2010,haldane2004}.

Here we study a novel class of exotic superconductivity
which can be realized in doped Weyl metals, and more generally in
systems with topologically non-trivial Fermi surfaces.
In superconductivity with pairing between states on two disjoint Fermi surface
sheets with
opposite Chern numbers, the Cooper pair inherits a non-trivial Berry
structure from the underlying single-particle Fermi surfaces.
Consequently, the pairing gap functions develop nontrivial net
vorticities leading to topologically-stable gapless nodes on the Fermi surfaces.
These nodes also determine the interplay between the surface modes
due to the Weyl point of the band structure and those
arising from the Cooper pairing.
For Fermi surfaces with approximate spherical symmetries,
the pairing symmetry can be classified by the monopole harmonic
functions rather than ordinary spherical harmonic functions.

We consider a general 
3D electron system with a
pair of separated Fermi surfaces, denoted as FS$_\pm$, respectively,
carrying opposite Chern numbers $\pm C$.
The doped Weyl metal can be thought as a concrete example.
Let us start with a minimal description that only assumes the existence
of parity (inversion) symmetry but broken TR symmetry.
In this model, there are
two Weyl points located at $\pm \mathbf{K}_0$,
and are related to each other by parity (inversion)
and respectively surrounded by FS$_{\pm}$.
Furthermore, the parity 
ensures that opposite monopole charges
$\pm q$ are enclosed by FS$_{\pm}$.
Define the electron creation operator $c^\dagger_a(\mathbf{k})$
in which $a$ is the index of a general two-band structure.
For the single-electron states on FS$_{\pm}$,
their creation operators are defined, respectively, as
$
\alpha_{\pm}^\dagger(\pm\mathbf{k})
=
\sum_{a}
\xi_{\pm,a}(\pm \mathbf{k}) c^\dagger_a(\pm \mathbf{K}_0 \pm \mathbf{k})
$,
in which $\pm \mathbf{k}$ are the relative momenta for electrons on
FS$_\pm$ with respect to $\pm \mathbf{K}_0$.
$\xi_{\pm,a}(\pm \mathbf{k})$ are the corresponding normalized eigen-functions
on FS$_\pm$, respectively.
And $\pm \mathbf{k}$ lie on two surfaces denoted as S$_\pm$ which
correspond to shifting FS$_\pm$ by $\mp \mathbf{K}_0$ towards the origin.
Because of the non-trivial monopole structure, $\xi_{\pm,a}(\pm\mathbf{k})$
cannot be globally well-defined for $\pm\mathbf{k}$ over the entire
surfaces of S$_\pm$, respectively.
They need to be described using a specific gauge.

The single-particle Berry connection can be defined as
$\mathbf{A}_\pm (\mathbf{k})= \sum_a
\xi_{\pm,a}^*(\mathbf{k}) i \mathbf{\nabla}_k \xi_{\pm,a} (\mathbf{k})$,
in which $\mathbf{\nabla}_k$ lies in the tangent space
of S$_\pm$, and, $\mathbf{A}_\pm$ is a tangent vector field therein.
The Berry fluxes satisfy
$\oiint_{S_\pm} d \mathbf{k} \cdot \nabla_k \times \mathbf{A}_\pm (\mathbf{k})
=\pm 4\pi q$.
The simplest case of $C=1$ is associated with the fundamental
monopole charge of $q=\frac{1}{2}$.

Let us consider the zero-momentum inter-Fermi surface pairing between FS$_+$ and FS$_-$. 
The pairing operator 
$P^\dagger(\mathbf{k})= \alpha^\dagger_+(\mathbf{k} )\alpha^\dagger_-(-\mathbf{k})$.
As has been pointed out by Murakami and Nagaosa in Ref. [\onlinecite{murakami2003a}],
the Berry connection of the two-particle state created by
$P^\dagger(\mathbf{k})$ 
is calculated as
$\mathbf{A}_p(\mathbf{k})=\mathbf{A}_+(\mathbf{k}) - \mathbf{A}_-(-\mathbf{k})$.
The total pairing Berry flux penetrating S$_+$ is
$
\oiint_{S_+} d\mathbf{k} \cdot \nabla_{k} \times \mathbf{A}_p(\mathbf{k})= 4\pi q_p
$
with $q_p=2q$.
In other words, the inter-Fermi surface Cooper pairing inherits the
Berry fluxes of two single-electron Fermi surfaces.
Consequently, the Cooper pairing phases cannot be well-defined
over the entire Fermi surfaces, which leads to generic nodal structure
of pairing gap functions.

Let us consider the gap function over  S$_+$ as $\Delta(\mathbf{k})$, which is
conjugate to the pairing
operator $P^\dagger(\mathbf{k})$ and is a single-valued complex
function.
Assuming the nodal structure of $\Delta(\mathbf{k})$ 
only composed of isolated points or lines, 
it can be proved that
$\Delta(\mathbf{k})$ possesses generic nodal structure
with the total vorticity $2q_p$, which
is a consequence of the band topology on FS$_{\pm}$
and is independent of specific pairing mechanisms and symmetries. 
The gap function $\Delta(\mathbf{k})$ can be parameterized as $|\Delta(\mathbf{k})|
e^{i\phi(\mathbf{k})}$ in which $\phi(\mathbf{k})$ is the pairing phase.
$\Delta(\mathbf{k})$ is gauge-covariant as follows:
Under the gauge transformation
$\xi_\pm (\pm \mathbf{k})\rightarrow \xi_\pm (\pm \mathbf{k}) e^{i\Lambda_\pm (\pm\mathbf{k})}$,
we have $\alpha_\pm^\dagger(\pm\mathbf{k})\rightarrow \alpha_\pm^\dagger(\pm\mathbf{k})
e^{i\Lambda_{\pm}(\pm\mathbf{k})}$, and $P^\dagger(\mathbf{k})\rightarrow
P^\dagger(\mathbf{k})e^{i\Lambda(\mathbf{k})}$ in which
$\Lambda(\mathbf{k})=\Lambda_+(\mathbf{k})+\Lambda_-(-\mathbf{k})$.
Consequently, $\phi(\mathbf{k})$ and $\mathbf{A}_p(\mathbf{k})$ transform
as $\phi(\mathbf{k})\rightarrow \phi(\mathbf{k})-\Lambda(\mathbf{k} )$,
and $\mathbf{A}_p(\mathbf{k})\rightarrow \mathbf{A}_p(\mathbf{k})-\mathbf{\nabla}_k \Lambda(\mathbf{k})$.
We define a gauge invariant $\mathbf{k}$-space circulation field 
on S$_+$ as
$\mathbf v(\mathbf{k}) = \mathbf{\nabla}_k \phi(\mathbf{k}) - \mathbf{A}_p$,
which is regular except at gap nodes.
If we consider the case that $\Delta(\mathbf{k})$ only has isolated
zeros located at $\mathbf{k}_{i} ~(i=1,2,..., n)$.
An infinitesimal oriented loop $C_i$ is defined around each zero $\mathbf{k}_i$
whose positive direction depends on the local normal direction by the right-hand rule.
Then, $\oint_{C_i} d\mathbf{k} \cdot \mathbf v=2\pi g_i$ in which
$g_i$ is the vorticity and integer-valued.
Next, reversing the direction of each loop $C_i$ and applying Stokes' theorem on
S$_+$ (excluding the bad points $\mathbf{k}_i$'s on which $\mathbf v$ is ill-defined),
we arrive at
\bea
\sum_i g_i=\frac{1}{2\pi} \sum_i \oint_{C_i} d \mathbf{k} \cdot \mathbf v =
\oiint \frac{d \mathbf{k}}{2\pi}  \cdot (\mathbf{\nabla}_k \times \mathbf{A}_p)=2 q_p. \nn
\\
\label{eq:vorticity}
\eea
This proof is gauge-independent.
If $\Delta (\mathbf{k})$ has line-nodes on S$_+$ which behave
as branch-cuts of $\mathbf v$, the proof can also be done
similarly. 

Consequently, when $q_p\neq 0$, 
$\Delta(\mathbf{k})$ cannot be a regular function
over the entire 
S$_+$.
Its nodal structure is distinct from that of the usual 
pairing symmetries
characterized by
spherical harmonics $Y_{lm}(\hat{\mathbf{k}})$, which are 
regular functions over the sphere.
The absence of the monopole structure gives rise to vanishing total
vorticity of phases.
For example, for the $^3$He-A type pairing with the orbital symmetry
$Y_{11}(\hat{\mathbf{k}})$, two gap nodes lie at the north and south poles
as a pair of vortex and anti-vortex of the pairing phase field,
respectively.

\begin{figure}[tbp]
\centering
\epsfig{file=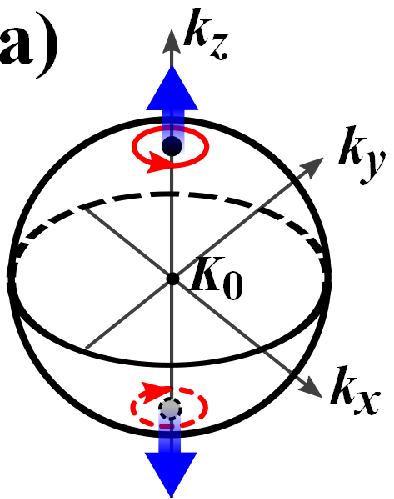,clip=1,width=0.3\linewidth,angle=0} \ \ \ \ \ \ \  \ \ \
\epsfig{file=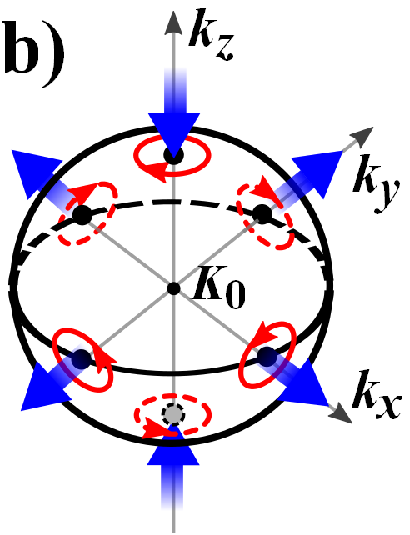,clip=1,width=0.3\linewidth, angle=0}
\caption{The vorticity and nodal structure of the pairing
gap function $\Delta(\mathbf{k})$ on the bulk Fermi surface $S_+$.
The total vorticities always equal $+2$. 
The vortices at the north and south poles have the vorticity
$\pm 1$ for ($a$) $\Delta_y=-i\Delta_x$ and ($b$) $\Delta_y=i\Delta_x$,
respectively.
Each of the four additional nodes in ($b$) exhibit the vorticity
$+1$.
}
\label{fig:vorticity}
\end{figure}

To illustrate this, we use a concrete simple model for
the 3D Weyl metal defined in a bipartite  array of lattice planes  with spinless
fermions \cite{hosur2012,haldane2014}.
\bea
H_K&=& \sum_{a,b} \sum_{\mathbf{k}}c^\dagger_a(\mathbf{k}) \Big\{
[t_- \cos(2k_x)+t_+(k_y,k_z)]\sigma_x +\nn \\
&+& \sin (2k_x) \sigma_y + V(k_y) \sigma_z -\mu I  \Big\}_{ab} c_b(\mathbf{k})
+h.c.,
\label{eq:weyl1}
\eea
in which the $\sigma_z$-eigenbasis refer to $A$ and $B$
sublattices; $V_{k_y}=2k_y$, $t_+(k_y,k_z)=-(k^2_y+k_z^2)$
and $t_-=1$.
$H_K$ is invariant under the inversion transformation with respect
to the center of a bond along $x$-direction, {\it i.e.},
$A\leftrightarrow B$, $k_{y,z} \rightarrow -k_{y,z}$, and
it breaks TR symmetry, because the ``spinless'' fermions can be
considered as fully-spin-polarized.  It also has a mirror symmetry $k_z\rightarrow -k_z$,
and has  a pair of Weyl points  at
$\pm \mathbf{K}_0=(0,0,\pm 1)$ on  the $k_z$-axis.
If $\mu>0$, the Fermi surface sheets FS$_\pm$ enclosing $\pm \mathbf{K}_0$
have  Chern numbers $C=\pm 1$, respectively.
For small values of $\mu$, FS$_\pm$ are approximately spherical.

Consider the following pairing Hamiltonian 
\bea
H_\Delta(\mathbf{k})&=&\sum_{a,b} \sum_{\mathbf{k}} c_a^\dagger(\mathbf{k})
[ 2i \Delta_x \sin (2k_x) I + \nn \\
&& \ \ \ \ +2 i \Delta_y \sin k_y\sigma _1]_{ab}  c_b^\dagger(-\mathbf{k}) + h.c.
\label{eq:pair}
\eea
The constructions of Eqs. (\ref{eq:weyl1}) and (\ref{eq:pair}) on the lattice
are presented in the Supplemental Material (Suppl. Mat.) A.
Assuming that the system is in the weak pairing region, {\it i.e.},
$|\Delta_{x,y}|\ll |\mu|$, we can project the pairing onto FS$_\pm$.
Since the gap function satisfies 
$\Delta(-\mathbf{k})=-\Delta(\mathbf{k})$,
we only need to consider $\Delta(\mathbf{k})$ on FS$_+$.
$\Delta(\mathbf{k})$ exhibits two nodes at the north and the south poles.
We choose two different gauges, which are non-singular in the
two polar regions.
Then, the gap function can be approximated as  $\Delta(\mathbf{k})=
\mp \Delta_x k_x + \Delta_y k_y $ at north and south poles,
respectively.
If $\Delta_y=-i\Delta_x$, the velocity field $\mathbf v(\mathbf{k})$
exhibits a pair of vortices located at north and south poles
with the vorticity $+1$ 
(Fig. \ref{fig:vorticity} (a)).
There are no other nodes on 
FS$_+$, and the total vorticity
is $+2$ in agreement with Eq. (\ref{eq:vorticity}). 
In contrast, if $\Delta_y=i\Delta_x$, each of the nodes at the north and
south poles changes its vorticity to $-1$.
To maintain the total vorticity, four additional nodes appear
on FS$_+$, each of which 
has vorticity 
$+1$
(Fig. \ref{fig:vorticity} (b)).
In our model, 
these four nodes are located on the equator with azimuthal angles
$\pm\frac{\pi}{4}, \pm\frac{3\pi}{4}$.

\begin{figure}[tbp]
\centering
\epsfig{file=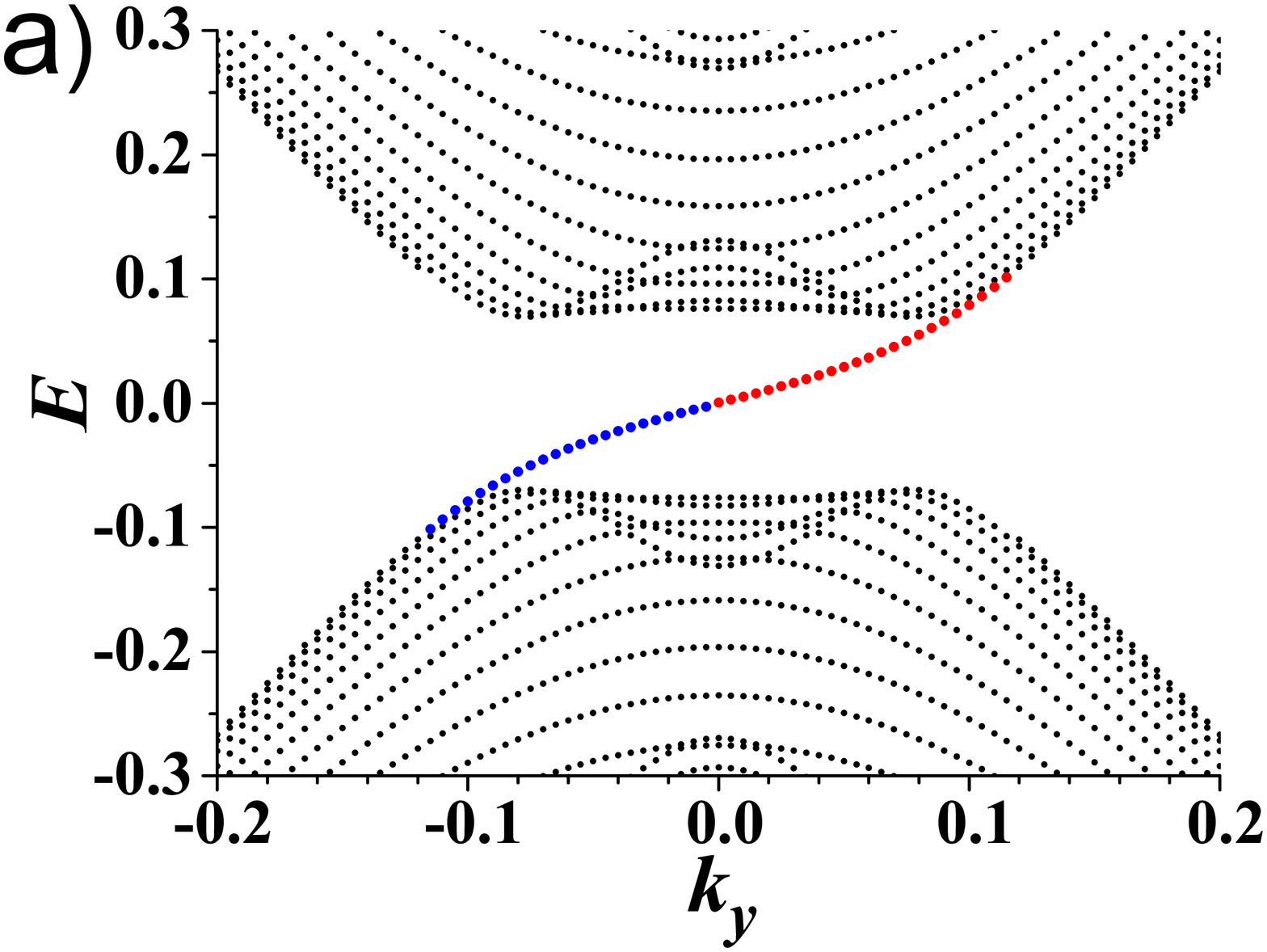,clip=1,width=0.49\linewidth, angle=0}
\epsfig{file=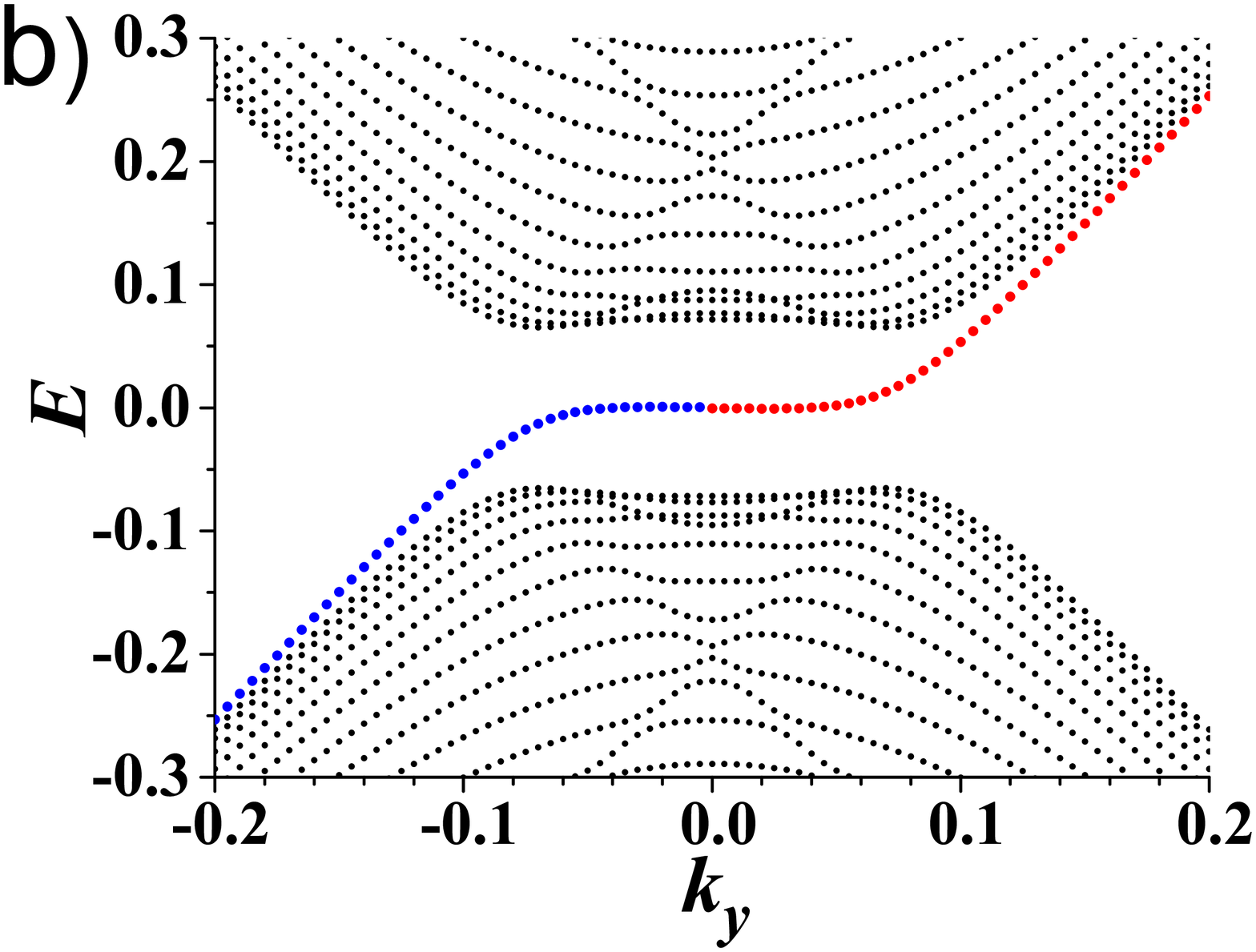,clip=1,width=0.49\linewidth, angle=0}
\epsfig{file=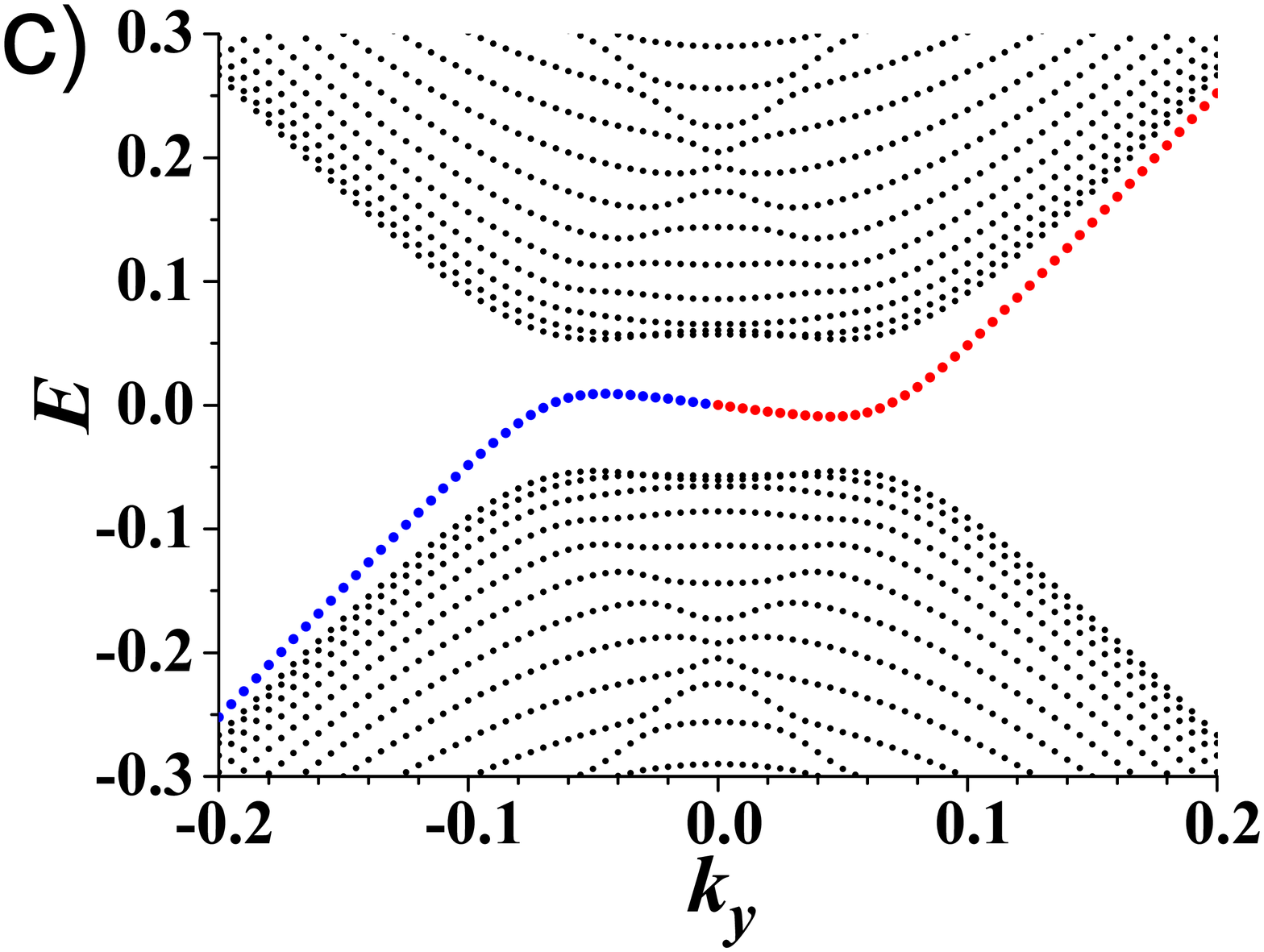,clip=1,width=0.49\linewidth, angle=0}
\epsfig{file=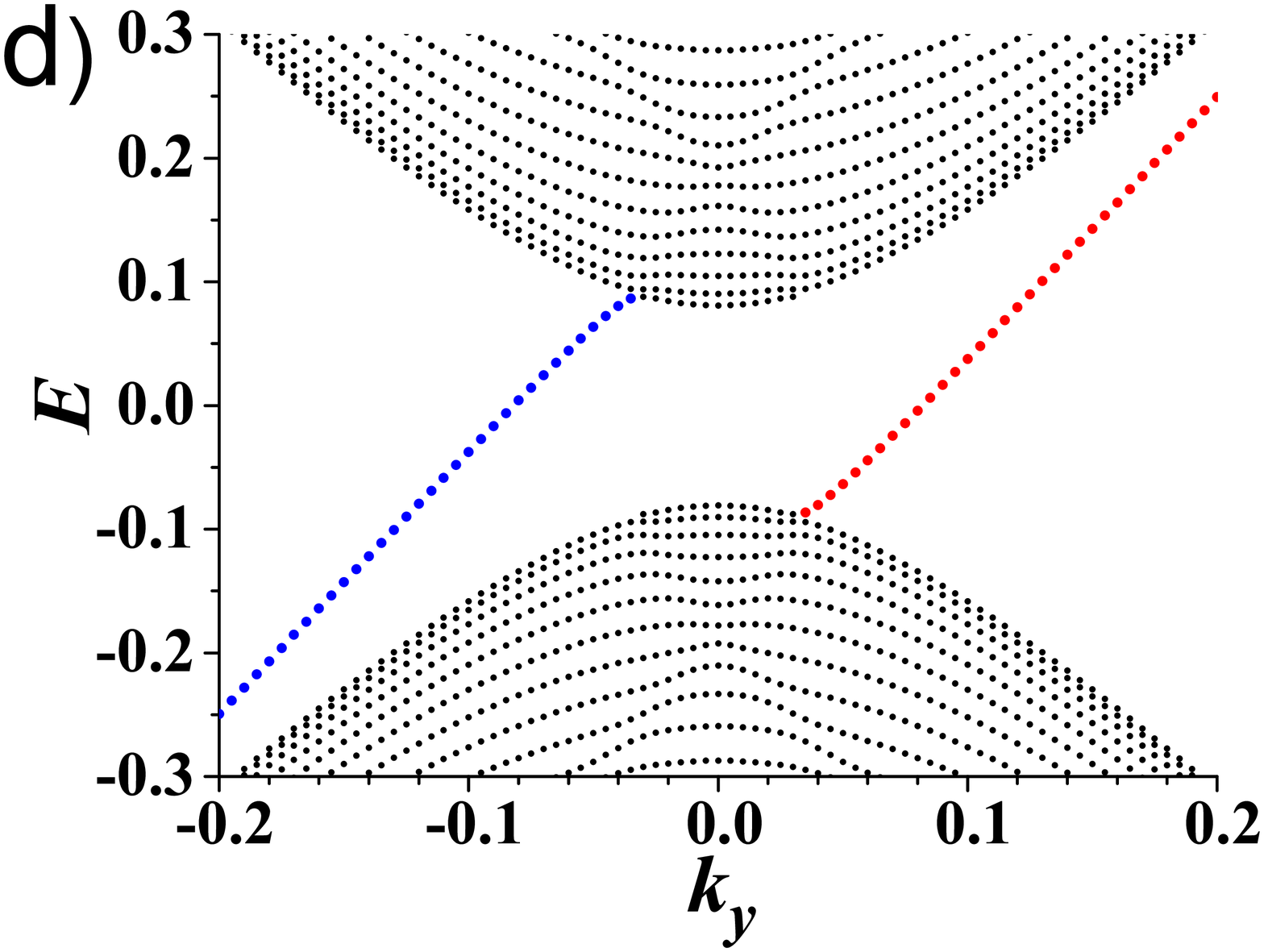,clip=1,width=0.49\linewidth, angle=0}
\caption{(Color Online) The projected bulk and surface mode spectrum on a
  single open boundary of the $yz$-plane
vs. $k_y$ at different cuts of $k_z=1.03, ~0.95, ~0.93, ~0.85$ from $a$)
to $d$), respectively.
The parameters are $\Delta_y=-i\Delta_x$ with $\Delta_x=0.2$ and $\mu=0.2$.
The cuts of $k_z$ lie between the intersecting points of the $z$-axis on FS$_+$ for ($a$), ($b$), and ($c$),
while $k_z$ in ($d$) lies outside FS$_+$.
The red (blue) color represents the positive (negative) charge carried by the surface modes.
}
\label{fig:surface_1}
\end{figure}

We also calculated the surface 
spectra on the open boundary.
In the absence of pairing, this model shows 
chiral surface states
for $k_z$ lying in the region
$-K_{0,z} <k_z< K_{0,z}$ 
\cite{haldane2014}.
Now, Cooper pairing opens pairing gaps on FS$_\pm$, and generates
additional Majorana surface modes inside the pairing gaps.
These Majorana surface modes are determined by the pairing nodal
structure on FS$_\pm$ and the associated vorticity pattern.
As have 
been described 
in Ref. [\onlinecite{lu2015}], these must
connect to the Fermi arcs
arising from the Weyl band structures as  $k_z$
varies.
\begin{figure}[tbp]
\centering
\epsfig{file=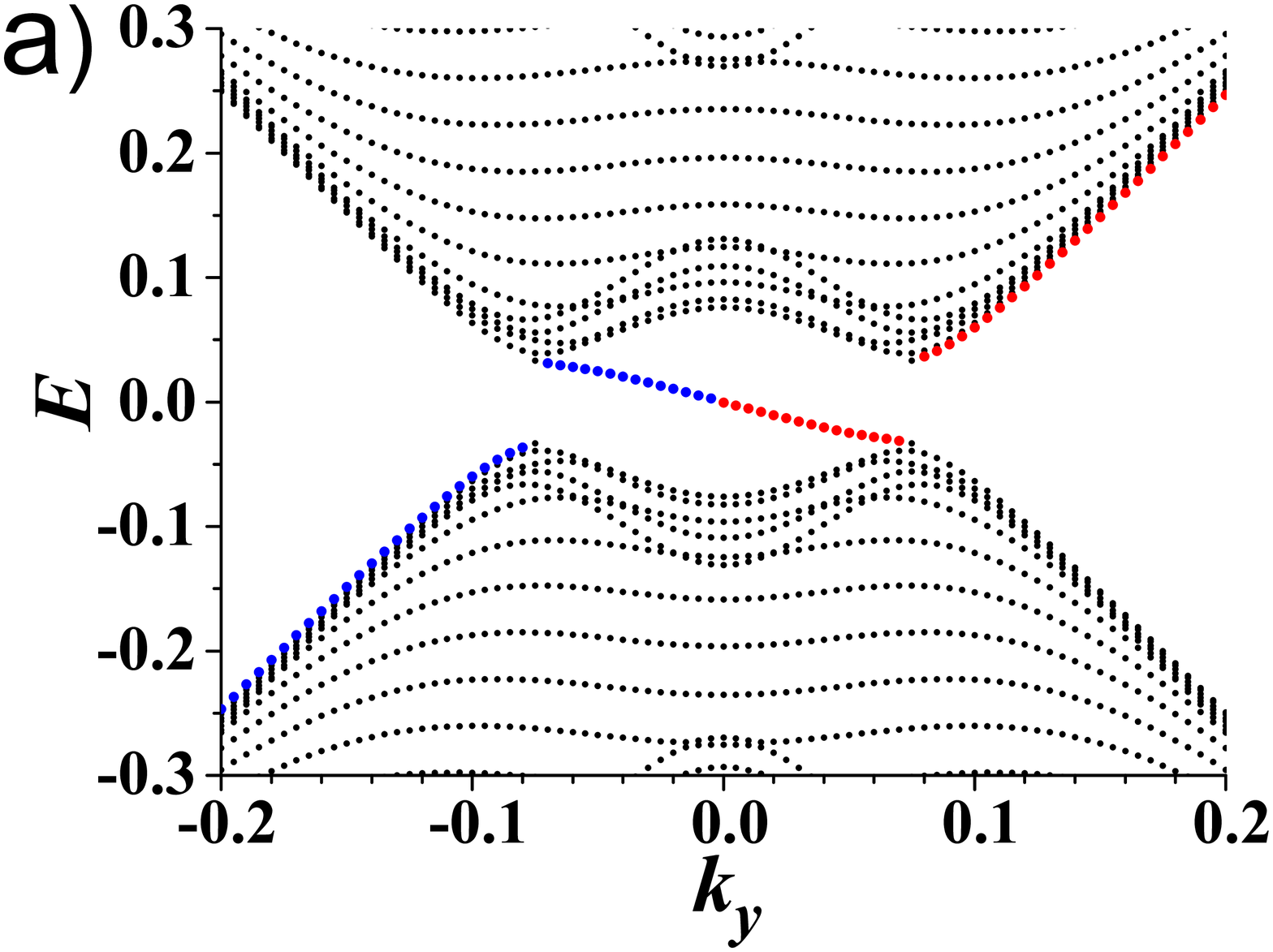,clip=1,width=0.48\linewidth, angle=0}
\epsfig{file=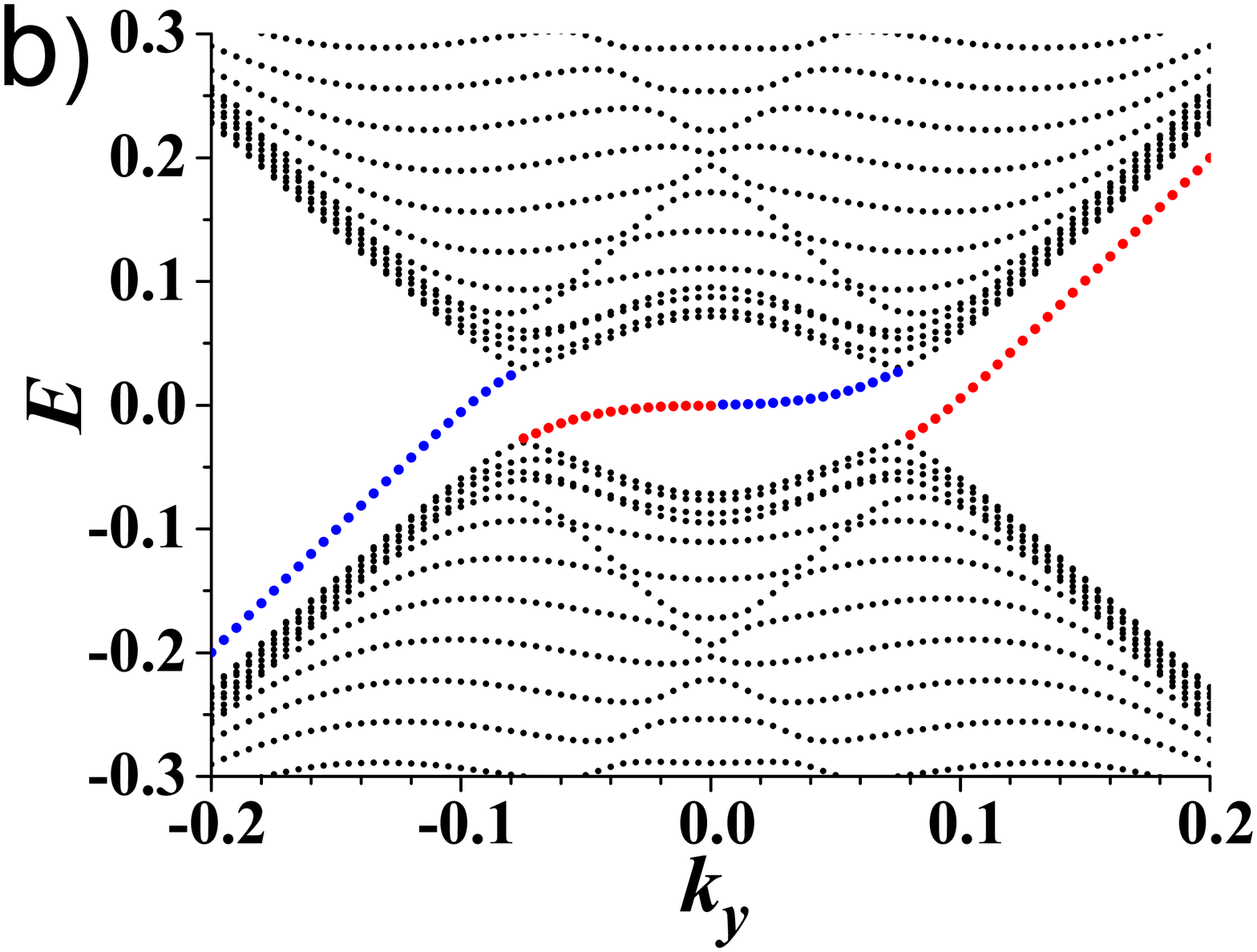,clip=1,width=0.48\linewidth, angle=0}
\epsfig{file=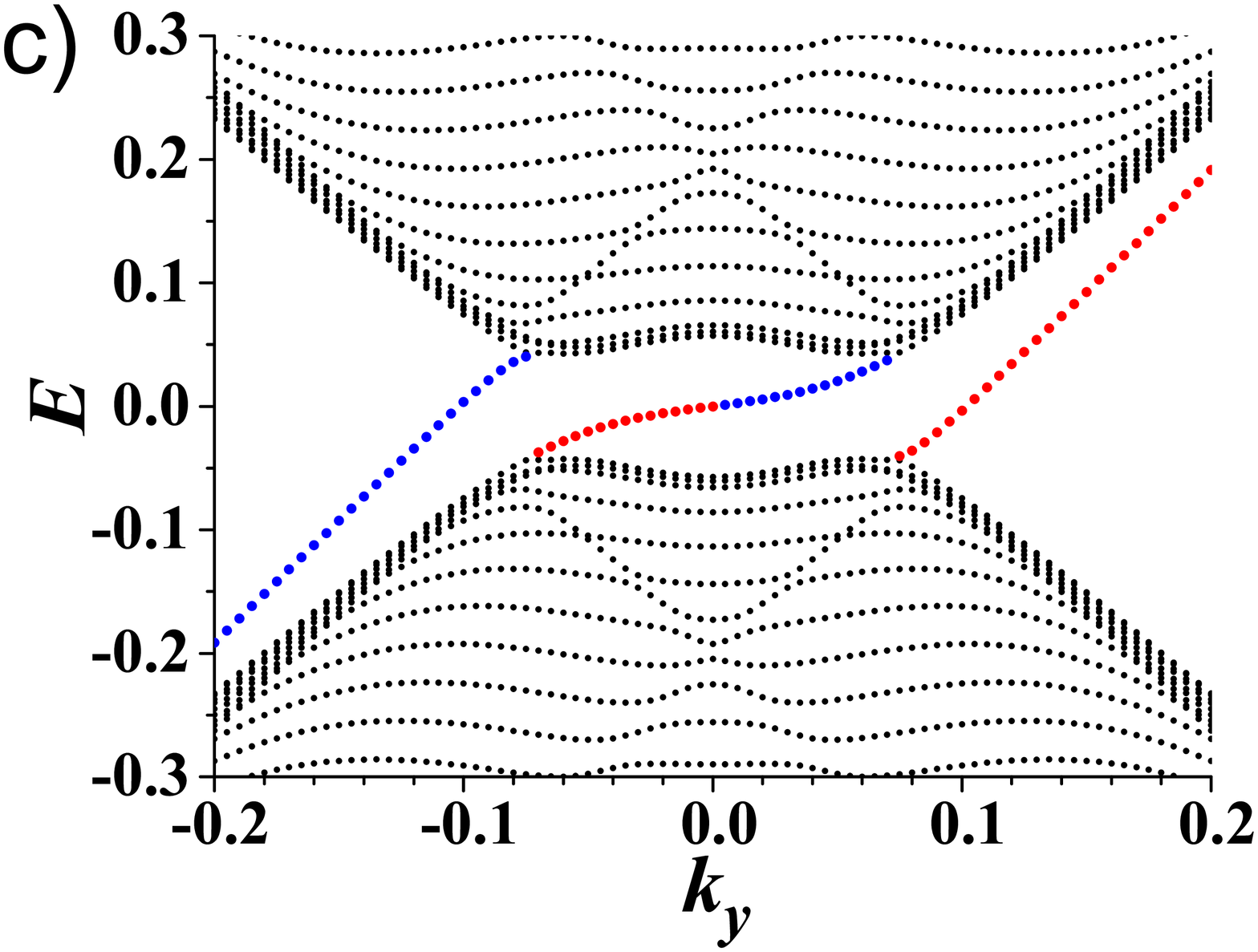,clip=1,width=0.48\linewidth, angle=0}
\epsfig{file=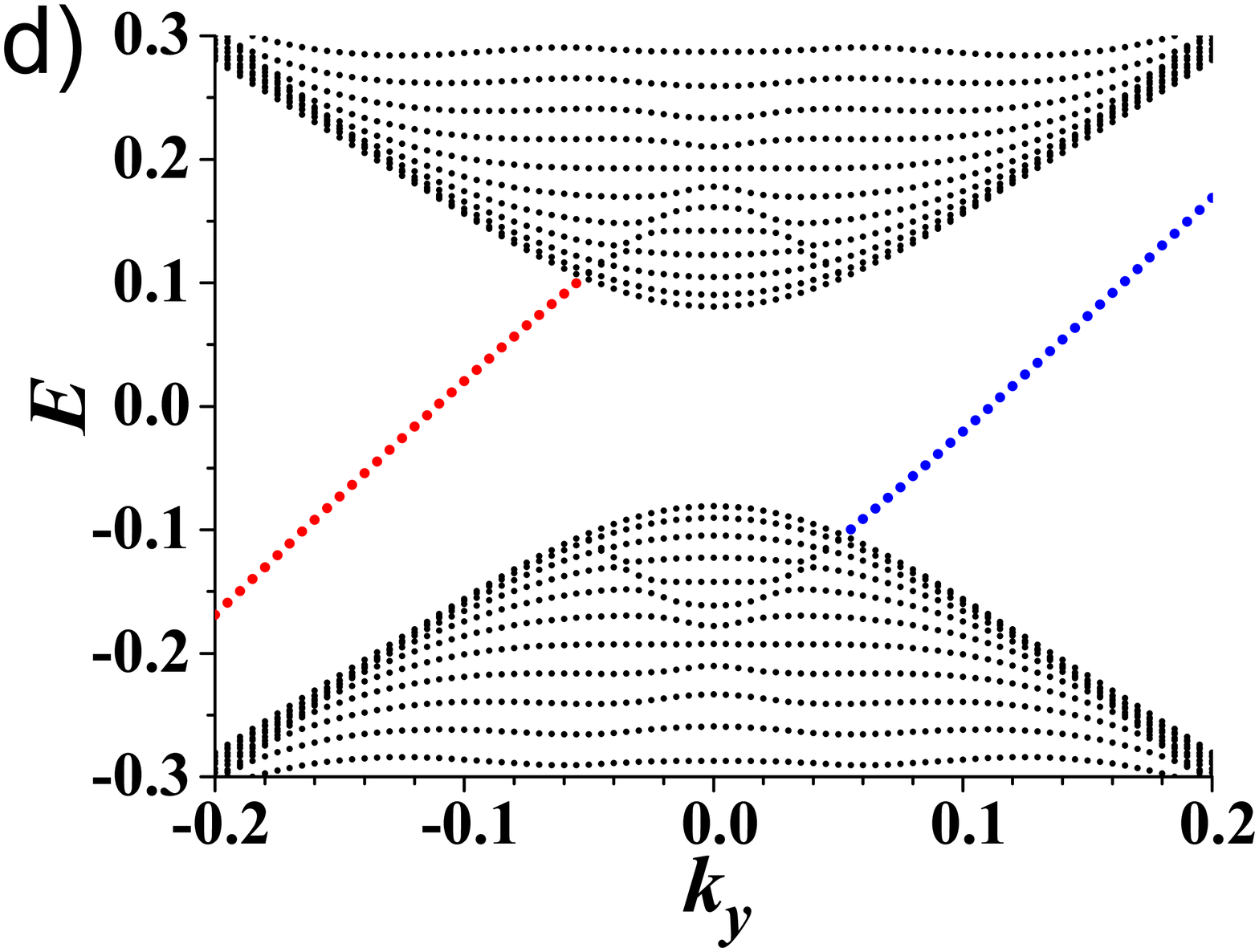,clip=1,width=0.48\linewidth, angle=0}
\caption{(Color Online) The bulk  and surface spectrum as in Fig. \ref{fig:surface_1},
except that $\Delta_y=i\Delta_x$ here.
The chiralities of the Majorana surface modes at $k_z$ close
to the north and south poles are opposite to those
in the case of $\Delta_y=-i\Delta_x$.
(A topological change  occurs at $k_z$ = 1.0, between (a) and (b).)}
\label{fig:surface_2}
\end{figure}

We impose two open boundaries parallel to the $yz$-plane,
and plot the spectra vs. $k_y$ for 
different 
$k_z$, with modes
localized on the bottom boundary suppressed.
The results of the case $\Delta_y=-i\Delta_x$ are shown
in Fig. \ref{fig:surface_1} ($a$)-($d$).
Under the Bogoliubov-de Gennes (BdG) formalism, there are four
quasiparticle bands,
but only states with $k_z>0$ are independent.
Because of the mirror symmetry, the spectrum is invariant under $k_z \mapsto -k_z$
plus a particle-hole transformation.
At $\mu=0.2$, FS$_+$ enclosing $K_0=(0,0,1)$ intersects the $k_z$-axis at $k_n\approx 1.1$
(the north pole) and $k_s\approx 0.9$ (the south pole).
For the cut of $k_z>k_n$, or, $k_z<k_s$, it does not intersect FS$_+$.
The corresponding surface spectra are determined by the Weyl
band structure: No surface modes exist at $k_z>k_n$, and
two branches of chiral surface modes appear at $-k_s<k_z<k_s$
(Fig. \ref{fig:surface_1} ($d$)).
They are related by particle-hole transformation
under which $(k_y,k_z)$ $\mapsto$ $(-k_y,-k_z)$, which reverses their
charge, followed by
$z$-reflection, $(-k_y,-k_z)$ $\mapsto$ $(-k_y,k_z)$,
Consequently, for fixed $-k_s < k_z  < k_s$ one surface mode is
entirely electron-like (the standard Fermi arc) with a quasiparticle charge $0 <
e^*(k_y,k_z) < e$, and the other is the $z$-reflection of its  particle-hole conjugate, with
$e^*(-k_y,k_z) < 0$.
Because $k_z$ lies outside FS$_\pm$, the particle-hole mixing is weak,
so the charge of the electron-line arc mode is close to $e$.

The  change in surface  band  topology  between $k_z>k_n$ and  $-k_s<k_z<k_s$
must occur through gap closings at $k_z$ = $k_n$ and $k_s$.
In this region, each $k_z$ defines a  Fermi-surface cross section
(``Fermi-CS'')  on
FS$_+$  of the Weyl metal,  which becomes  gapped by pairing.
The only surviving bulk zero-energy excitations at the nodal points
 are 
Weyl-Majorana (WM) points in the BdG formalism,
and classified as positive or negative according to their
chiral indices.
The two WM points at the north and south poles both
carry positive pairing vorticities $+1$.
As $k_z$  decreases through the WM point at the north pole,
the surface gap closes and reopens with a single surface mode
passing through zero energy,
as shown in Fig. \ref{fig:surface_1}($a$-$c$).
After $k_z$ passes the WM point at the south pole, which also has
pairing vorticity $+1$,
the number of the branches of surface modes is increased to 2, as in
the normal Weyl metal.

\begin{figure}[tbp]
\centering
\epsfig{file=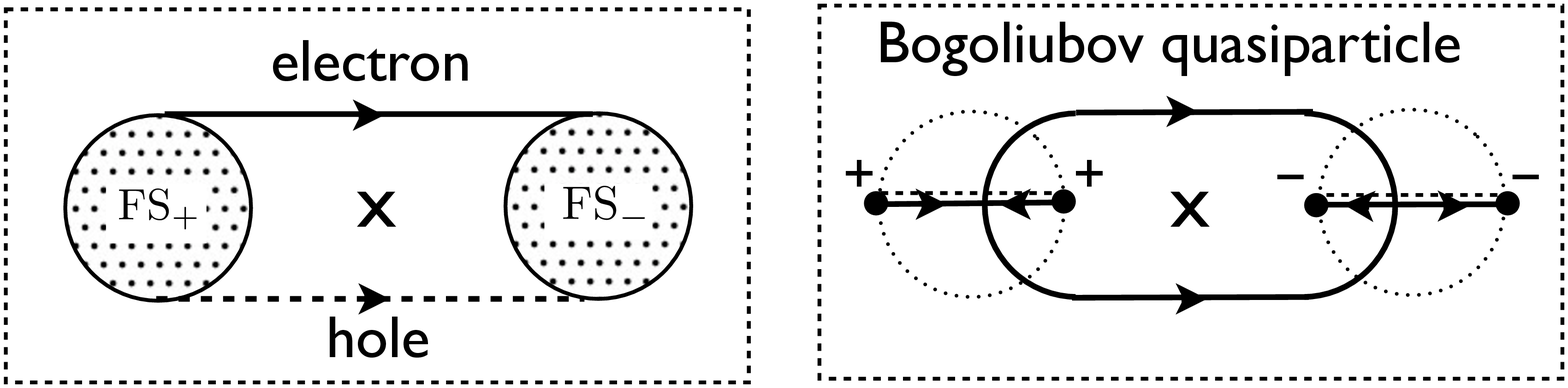,clip=1,width=0.9\linewidth, angle=0}
\caption{(Schematic.)
Zero-energy surface states in the surface Brillouin zone, with a
chiral pair of projected
bulk Fermi surfaces (left)
or four projected bulk gapless superconductor nodes (right); ``$X$'' marks
$\mathbf k$ = 0.
The directions on arcs are $\hat {\mathbf{n}} \times
\hat{\mathbf{v}} $, where $\hat {\mathbf{n}}$ is the normal to the
surface, and $\mathbf{v}$ is the group velocity of the surface mode.
The ``flow"\cite{haldane2014,lu2015} starts and ends at projections
of sources and sinks of 3D Berry flux at the Fermi level.  Right: a
``vector central charge" $\tilde c \hat{\mathbf{v}}$ with $\tilde c$ =
$\frac{1}{2}$ is associated with the Bogoliubov edge modes (directed
solid lines), with $\sum \tilde c \hat{\mathbf{v}} =0$ at the
``cross'', which is a special feature\cite{lu2015} of  mirror
symmetry.   Also shown (dashed line connecting nodes) is the
quasiparticle ``neutrality line'', which is only pinned to zero energy by
the mirror symmetry.
}
\label{fig:locus}
\end{figure}

When $k_n>k_z>k_s$, the surface quasiparticle charge  changes continuously
as a function of $k_y$ from hole-like to particle-like at a ``neutral
point'', which in our model is pinned at $k_y$ = 0 by
the $z$-reflection symmetry $e^*(k_y,k_z)$ $\equiv$ $-e^*(-k_y,-k_z)$
= $e^*(-k_y,k_z)$.
In general, these points  are on  a ``neutrality line'' in the surface
BZ  connecting the projections of the two
WM points.   The $z$-reflection symmetry also gives the quasiparticle
spectrum the symmetry $E(k_y,k_z)$ $\equiv$ $-E(-k_y, -k_z)$ =
$E(-k_y,k_z)$, so the zero-energy line, like the  neutral point is
pinned to $k_y$ = 0, and its group velocity is in the $y$ direction.
Near the north pole, the zero-energy point has group-velocity
in the $+\hat{\mathbf y}$ direction, while near the south
pole, it is along  $-\hat{\mathbf y}$.
A consequence of the
  reflection symmetry is that at  some intermediate $k_z$, its group velocity
  vanishes,  and for fixed $k_z$ less than this,  there are three values of $k_y$ with
  a zero-energy quasiparticle, one electron-like, one hole-like, and
  one neutral.     At this inflection point, the locus of zero-energy
lines in the surface BZ has a ``cross shape''\cite{lu2015} (see
Fig.\ref{fig:locus}),
that is a symmetry-protected feature of the  $z$-reflection symmetry.

The surface spectra at the case of $\Delta_y=i\Delta_x$ were
also calculated (Fig.\ref{fig:surface_2}).
The surface modes in the cases of $k_z>k_n$ and $k_s>k_z>0$
are not directly related to pairing, and thus are qualitatively
the same as the case of $\Delta_y=-i\Delta_x$.
However, the Majorana surface modes at $k_n>k_z>k_s$ are markedly
different due to the more complicated nodal structure on FS$_+$.
As shown in
Fig.\ref{fig:vorticity}($b$), the WM points at the north
and the south poles are negative; while those near
the equator are positive.
As $k_z$ is reduced below $k_n$, a Majorana surface mode appears
inside the pairing gap as in Fig.\ref{fig:surface_2}($a$), with
opposite  chirality to that in Fig.\ref{fig:surface_1}($a$).
The gap closes again at $k_z$ = 1, where  projections of the four
positive WM points are found.
When the gap reopens, there are three modes with positive chirality,
the neutral one of which disappears  when the gap closes at the
now-negative projected WM point at  $k_z$ = $k_s$.


Next we study the pairing partial-wave symmetries when FS$_\pm$ have
approximate spherical symmetry.
If we neglect the small anisotropy, the complete bases of $\Delta (\mathbf{k})$
for $\mathbf{k}$ lying on S$_+$ with the total vorticity $q_p$ can be spanned
by the monopole harmonic functions $Y_{q_p,jm}(\hat{\mathbf{k}})$ instead of the
usual $Y_{lm}$.
Monopole harmonic functions have been widely applied in physics
\cite{wu1976,wu1977,haldane1983}.
For completeness, their basic properties are summarized in
Suppl. Mat. B. 
After projecting the pairing Hamiltonian to FS$_\pm$,
it becomes $H_\Delta= \sum_{\mathbf{k}} \Delta
(\mathbf{k}) P^\dagger(\mathbf{k}) +\Delta^*(\mathbf{k}) P(\mathbf{k})$
for $\mathbf{k}$ lying close to S$_+$.
We define $\Delta (\mathbf{k})= \Delta(|k|) f(\hat{\mathbf{k}})$,
in which the angular dependence on $\hat{\mathbf{k}}$ and the energy dependence
on $|k|$ are separated.
$\Delta(|k|)$ is assumed positive and the angular factor $f(\hat
k)$ is complex satisfying the normalization condition $\int d\hat{\mathbf{k}} |f(\hat
k)|^2=1$.
$f(\hat{\mathbf{k}})$ is expanded in terms of the monopole harmonic functions as
\bea
f(\hat{\mathbf{k}}) = \sum_{jm} c_{jm} \mathcal{Y}_{q_p;jm} (\hat{\mathbf{k}}),
\label{eq:gap}
\eea
in which $c_{jm}$ are complex coefficients. Both the pairing operator $%
P^\dagger(\mathbf{k})$ and the gap function $\Delta(\mathbf{k})$ are gauge
dependent, while, $H_\Delta$ is gauge independent.

A remarkable feature is that all the pairing channels should
have $j\ge |q_p|$ regardless of the pairing mechanism
since $\mathcal{Y}_{q_p,j,m}$ starts with $j=|q_p|$.
The absence of pairing channels with $j<|q_p|$ is robust, as 
a consequence of topology and the monopole harmonic representation
of the rotation group.
Furthermore, the lowest order pairing channel $j=|q_p|$ is special:
$\mathcal{Y}_{q_p,j=|q_p|,m} (\hat{\mathbf{k}})$ are holomorphic or anti-holomorphic
functions.
All of its $2q_p$-nodes exhibit the same vorticity, and thus
$\Delta(\mathbf{k})$ is completely determined by the
locations of its nodes up to an overall factor.
The nodes of $\mathcal{Y}_{q_p,j=|q_p|,m} (\hat{\mathbf{k}})$ represent vortices
of the pairing phases on S$_+$.
The location of pairing nodes are also WM 
points of the BdG Hamiltonian with the same chirality.
For each node on FS$_+$ with $\mathbf{K}_0+\mathbf{k}$, there exists its
image on FS$_-$ exhibiting the opposite vorticity.

Let us consider a concrete example of spin-$1/2$ fermions 
in the continuum with spin-orbit coupling.
The low energy kinetic Hamiltonians around the Weyl points are
\bea
H_{\text{Weyl},\pm}(\pm\mathbf{K}_0+\mathbf{k})= \pm v_F \mathbf{\sigma} \cdot \mathbf{k} -
\mu,
\label{eq:weyl}
\end{eqnarray}
where $\sigma$'s represent Pauli matrices for spins;
$\mu>0$ is assumed without loss of generality.
We choose the  gauge convention:
$\xi_{+}(\hat{\mathbf{k}})= \xi_{-}(-\hat{\mathbf{k}})= \left(
u_{\hat{\mathbf{k}}}, v_{\hat{\mathbf{k}}}\right)^T$,
in which $u_{\hat{\mathbf{k}}}=\cos\frac{\theta_{\hat{\mathbf{k}}}}{2}$ and $v_{\hat{\mathbf{k}}}=\sin%
\frac{\theta_{\hat{\mathbf{k}}}}{2} e^{i\phi_{\hat{\mathbf{k}}}}$, which
are singular when $\hat{\mathbf{k}}$ is located at the north and
the south poles on FS$_\pm$, respectively.
$P^\dagger(\mathbf{k})=\alpha_+^\dagger(\mathbf{k}) \alpha_-^\dagger(-\mathbf{k})$
is a spin-$1$ helicity eigen-operator, satisfying
$[\mathbf{S} \cdot \hat{\mathbf{k}}, P^\dagger (\mathbf{k})] = P^\dagger (\mathbf{k})$,
in which $\mathbf{S}$ is the total spin of the Cooper pair.
Under the $S_z$-eigenbasis,
$P^\dagger(\mathbf{k}) =\sum_{m=-1}^{1}\sqrt{\frac{4\pi}{3}}
\mathcal{Y}^*_{-1;1m} (\hat{\mathbf{k}}) \chi^\dagger_{1m} (\mathbf{k})$,
in which $\chi^\dagger_{1m}(\mathbf{k})=\avg{1m|\frac{1}{2}\sigma
\frac{1}{2}\sigma^\prime}
c^\dagger_\sigma(\mathbf{K}_0+\mathbf{k})c^\dagger_{\sigma^\prime}(-\mathbf{K}_0-\mathbf{k})$.
The monopole charge enclosed by S$_+$ is $q_p=-1$
for the phase distribution of $\Delta(\mathbf{k})$.
According to Eq. (\ref{eq:gap}), by setting $j=|q_p|=1$,
$f(\hat{\mathbf{k}})$ is a quadratic homogeneous
function of $u_k$ and $v_k^*$, $
f(\hat{\mathbf{k}})=f(u_k,v_k)=\frac{1}{\sqrt N} \prod_{i=1,2} (u_k v_i^*-u_i v_k^*)$,
where $N$ is the normalization factor; the nodal points
are represented by
$\left(u_{i},v_{i}\right)^T
=\left(
\cos\frac{\theta_{i}}{2},
\sin\frac{\theta_{i}}{2} e^{i\phi_i}
\right)^T$ with $i=1,2$.

The locations of nodes in $f(\hat{\mathbf{k}} )$ are determined by the energetics.
The Ginzburg-Landau (GL) analysis 
(see Suppl. Mat. C.) 
shows that there are two typical possibilities for $j=1$ that $m$ in Eq. (\ref{eq:gap})
equals 0, or $\pm 1$:
In the former case,
\bea
\Delta_{m=0}(\mathbf{k})= \Delta(|k|) \sin\theta_k e^{i\phi_k}.
\label{eq:polar}
\eea
In contrast to the usual case where $\Delta(\mathbf
k)\propto Y_{10}(\hat{\mathbf{k}})$ exhibits a nodal line,
Eq. (\ref{eq:polar}) only has nodal points that
repel each other, at antipodal points on  S$_+$.
In realistic systems, the spherical symmetry of S$_+$ is broken
down to a lower point-group symmetry.
Since the pairing gap is suppressed around gap nodes,
the gap nodes may preferentially locate at the minima of local density
of states on S$_+$.
In the lattice case, two nodes attract each other and merge into a double
one.
Without loss of generality, they can locate at the north or
south pole as
\begin{eqnarray}
\Delta(\mathbf{k}) = \left\{
\begin{array}{l}
\Delta(|k|)\cos^2\frac{\theta_k}{2} ~~~~~~~~(m=~~1), \\
\Delta(|k|)\sin^2\frac{\theta_k}{2} e^{2i\phi_k} ~~(m=-1),%
\end{array}
\right.
\label{eq:axial}
\end{eqnarray}
respectively.
Again, its nodal structure is distinct 
from the
usual axial pairing 
with orbital symmetry characterized by spherical harmonic function
$\mathcal{Y}_{1\pm1}(\hat{\mathbf{k}})$.
The pairing structures of Eq. (\ref{eq:polar}) and Eq. (\ref{eq:axial})
in the representation of $\sigma_z$-basis
have also been studied in the context of pairing with magnetic
dipolar interactions \cite{li2012b}.

The monopole Cooper-pairing described above can be generalized to Fermi surfaces
carrying opposite Chern numbers $\pm C$ with $C\ge 2$. In these cases,
the monopole charge enclosed by the pairing surface $S_+$ equals $q_p=C$.
The mean-field free energy favors single nodes,
and nodes repel each other forming a vortex lattice configuration, 
although optimal configurations 
may be complicated by energetic issues.

We have also performed the partial-wave analysis of  the pairing
interactions 
(see Suppl. Mat. D) 
to show how the non-trivial topology of FS$_\pm$ transforms the ordinary
partial-wave channels into those characterized by monopole harmonics
starting with $j=|q_p|$.
Another possibility of Cooper pairing in doped Weyl metals is intra-Fermi
surface pairing with non-zero center of mass momenta, whose
pairing phase structure does not possess non-trivial Berry phase
structure on FS$_\pm$ (see  Suppl. Mat. E.).

In summary, we have studied the Cooper pairing structure
between two separate Fermi surfaces 
 carrying opposite Chern numbers $\pm C$.
The Cooper pairs carry a non-trivial Berry phase structure
characterized by the monopole charge $q_p=C$ so that their
phases cannot be globally well-defined on the Fermi surfaces.
The gap function $\Delta(\mathbf{k})$ generically possesses nodes with
the total vorticity $2q_p$.
These nodes are also the WM points of the Hamiltonian in the
BdG formalism.
The surface modes arise both from the Weyl band structure
and the pairing: the former exist inside the band gap,
while the latter appear inside the pairing gap on FS$_\pm$.
In a simplified model where FS$_{\pm}$ are both spherical, the pairing symmetry is
classified in terms of the monopole harmonic functions.
The lowest pairing channel is $j=|q_p|$ purely determined by
the symmetry rather than interaction, and the
corresponding pairing functions are holomorphic or anti-holomorphic
functions on the pairing surface.

\textit{Acknowledgments}
Y.L. thanks the Princeton Center for Theoretical Science at Princeton
University for support.
F.D.M.H. was supported in part by the MRSEC program at Princeton
Center for Complex Materials, grant NSF-DMR-1420541, and by the W. M. Keck Foundation.

\bibliographystyle{apsrev4-1}
\bibliography{all}

\newpage

\begin{center}{\textmd{\textbf{Supplemental Materials}}}
\end{center}

\subsection{Lattice construction of the Weyl metal model}
The Weyl metal model Eq. (2
) studied in the main text
can be formulated in the real space in a bipartite lattice with spinless
fermions \cite{haldane2014}.
After a partial Fourier transformation over the $y$ and $z$-directions,
the band Hamiltonian along the $x$-axis becomes
\bea
H_{k_y,k_z}^K&=& V_{k_y}\sum_i \Big\{ c^{A,\dagger}_{k_y,k_z}(i) c^A_{k_y,k_z}(i) -
c^{B,\dagger}_{k_y,k_z}(i) c^B_{k_y,k_z}(i) \Big\} \nn \\
&+& \sum_i \Big\{
t_{+}(k_y,k_z) c^{B,\dagger}_{k_y,k_z}(i) c^A_{k_y,k_z}(i+1) \nn \\
&+& t_- c^{A,\dagger}_{k_y,k_z} (i) c^B_{k_y,k_z}(i) +h.c.
\Big\},
\eea
in which $i$ is the index of unit cells containing $A$ and $B$ sites,
$V_{k_y}=2k_y$, $t_+(k_y,k_z)=-(k^2_y+k_z^2)$
and $t_-=1$.
$H^K$ is invariant under the inversion transformation with respect
to the center of a bond along $x$-direction, {\it i.e.},
$A\leftrightarrow B$, $k_{y,z} \rightarrow -k_{y,z}$, and
it breaks TR symmetry.
This Hamiltonian gives rise to a pair of Weyl point located at
$\pm \mathbf{K}_0=(0,0,\pm 1)$, respectively.
For chemical potential $\mu>0$, FS$_\pm$ enclosing $\pm \mathbf{K}_0$
possess the Chern number $C=\pm 1$, respectively.
Consider the following pairing structure
\bea
H^\Delta_{k_y,k_z}&=&\sum_{i,a=A,B}
\Big\{
\Delta_x c^{a,\dagger}_{k_y,k_z}(i) c^{a,\dagger}_{-k_y,-k_z}(i+1)+h.c.
\Big\}\nn \\
&+& \sum_i \Big\{ 2i \Delta_y \sin k_y
c^{A,\dagger}_{k_y,k_z}(i) c_{-k_y,-k_z}^{B\dagger}(i)
+h.c.\Big\}. \ \ \,  \ \ \,
\eea

After projection to FS$_+$, the gap function $\Delta(\mathbf{k})$
exhibits two nodes at the north and south poles, respectively.

\subsection{Monopole harmonic functions}
\label{sect:mono}

In this section, we present the definition and basic properties of monopole
harmonic functions.
For convenience, we will formulate mostly in real space, and
their momentum version can be obtained by replacing $\hat {\mathbf{r}}$
with $\hat{\mathbf{k}}$.

Consider a magnetic monopole located at the origin. For convenience, we
choose the following gauge to describe its vector potential $\mathbf{A}$ as
\begin{eqnarray}
\mathbf{A}(\mathbf r)=\frac{g}{|r|} \frac{\hat {\mathbf z}\times \mathbf
  r}{|r|+\mathbf r \cdot \hat {\mathbf z}}
=\frac{g}{r}\tan \frac{\theta}{2} \hat e_\phi,
\end{eqnarray}
which is singular along the Dirac string from the origin to the south pole.
The mechanical angular momentum is defined as $\mathbf \Lambda=\mathbf r \times
(\mathbf p -\frac{e}{c}\mathbf{A})$, but $\mathbf \Lambda$ does not satisfy the
commutation of the SU(2) algebra. The angular momentum satisfying the SU(2)
algebra is define as
\begin{eqnarray}
\mathbf L= \mathbf \Lambda -q \hat r,
\end{eqnarray}
where $q=\frac{eg}{c\hbar}$. $q$ is a positive integer, or,
half-integer, taking values of $\frac{1}{2},1, \frac{3}{2},2, ..$. Because $%
\mathbf \Lambda \perp \mathbf r$, we can verify that
\begin{eqnarray}
\Lambda^2 = L^2 -\hbar^2 q^2,
\end{eqnarray}
which is an operator identity. If an electron is confined on a sphere with
radius $R$ with a monopole of charge $q$ located in the center of the
sphere, its Hamiltonian can be described as
\begin{eqnarray}
H=\frac{\hbar^2}{2mR^2} \Lambda^2
\end{eqnarray}

The components of $L_{\pm}=L_x\pm iL_y$ and $L_z$ are
\begin{eqnarray}
L_z&=&-i\hbar \frac{\partial}{\partial \phi}-\hbar q  \notag \\
L_{\pm}&=&\hbar e^{\pm i \phi}(\pm \frac{\partial}{\partial \theta}
+i\cot\theta \frac{\partial}{\partial\phi}-q \tan\frac{\theta}{2}),
\end{eqnarray}
$L^2$ is expressed as
\begin{eqnarray}
\frac{L^2}{\hbar^2}&=&\frac{-1}{\sin^2\theta} (\sin\theta\frac{\partial}{%
\partial \theta}(\sin\theta\frac{\partial}{\partial \theta}  \notag \\
&+&\frac{1}{\sin^2\theta} (i\frac{\partial}{\partial \phi} +
q(1-\cos\theta))^2 + \hbar ^2 q^2.
\end{eqnarray}

The monopole harmonic functions $\mathcal{Y}_{q;jj_{z}}(\theta ,\phi )$ are
defined as
\begin{eqnarray}
L^{2}\mathcal{Y}_{q;jj_{z}}(\theta ,\phi ) &=&j(j+1)\hbar ^{2}\mathcal{Y}%
_{q;jj_{z}}(\theta ,\phi )  \notag \\
L_{z}\mathcal{Y}_{q;jj_{z}}(\theta ,\phi ) &=&j_{z}\hbar \mathcal{Y}%
_{q;jj_{z}}(\theta ,\phi ),
\end{eqnarray}%
where $j=|q|,|q|+1,.....$, and $j_{z}=-j,-j+1,...,j$. There is a nice
relation between the monopole harmonic functions and the rotation $D$-matrix
as
\begin{eqnarray}
\mathcal{Y}_{q;jm}(\theta ,\phi ) &=&\sqrt{\frac{2j+1}{4\pi }}\Big[%
D_{m,-q}^{j}(\phi ,\theta ,-\phi )\Big]^{\ast }  \notag \\
&=&\sqrt{\frac{2j+1}{4\pi }}e^{i(m+q)\phi }d_{m,-q}^{j}(\theta )
\label{eq:monoDmatrix}
\end{eqnarray}%
where $D_{m,m^{\prime }}^{j}(\alpha ,\beta ,\gamma )$ is defined in the
standard way as
\begin{eqnarray}
D_{m,m^{\prime }}^{j}(\alpha ,\beta ,\gamma ) &=&\langle jm|e^{-iJ_{z}\alpha
}e^{-iJ_{y}\beta }e^{-iJ_{z}\gamma }|jm^{\prime }\rangle  \notag \\
&=&e^{-im\alpha -im^{\prime }\gamma }d_{m,m^{\prime }}^{j}(\beta ),
\end{eqnarray}%
and $d_{m,m^{\prime }}^{j}(\beta )=\langle jm|e^{-iJ_{y}\beta }|jm^{\prime
}\rangle $. The expression for $d_{m,m^{\prime }}^{j}(\beta )$ is
\begin{eqnarray}
d_{m,m^{\prime }}^{j}(\beta ) &=&\sqrt{\frac{(j+m^{\prime })!(j-m^{\prime })!%
}{(j+m)!(j-m)!}}\big(\cos \frac{\beta }{2}\big)^{m^{\prime }+m}  \notag \\
&\times &\big(\sin \frac{\beta }{2}\big)^{m^{\prime }-m}P_{j-m^{\prime
}}^{m^{\prime }-m,m^{\prime }+m}(\cos \beta ),\ \ \,\ \ \,
\end{eqnarray}%
where the Jacobi polynomial $P_{j-m^{\prime }}^{m^{\prime }-m,m^{\prime
}+m}(\cos \beta )$ follows the definition as
\begin{eqnarray}
P_{n}^{a,b}(x) &=&\frac{(-)^{n}}{2^{n}n!}(1-x)^{-a}(1+x)^{-b}  \notag \\
&\times &\frac{d^{n}}{dx^{n}}[(1-x)^{a+n}(1+x)^{b+n}].
\end{eqnarray}

It will also be useful to introduce the two-component spinor notation $%
u=\cos \frac{\theta }{2},v=\sin \frac{\theta }{2}e^{i\phi }$. In this
notation, the expression of $\mathcal{Y}_{-q;j=q,j_{z}}(\theta ,\phi )$ is
simple in which $q>0$ as
\begin{eqnarray}
\mathcal{Y}_{-q;j=q,m}(\theta ,\phi ) &=&\left\{ \frac{2q+1}{4\pi }\left(
\begin{array}{c}
2q \\
q-m%
\end{array}%
\right) \right\} ^{\frac{1}{2}}u^{q+m}v^{\ast ,q-m}.  \notag \\
&&
\end{eqnarray}%
The relation between $\mathcal{Y}_{q;jm}(\theta ,\phi )$ and $\mathcal{Y}%
_{-q;jm}(\theta ,\phi )$ is
\begin{equation*}
\mathcal{Y}_{q;jm}(\theta ,\phi )=(-)^{q+m}\mathcal{Y}_{-q;j-m}^{\ast
}(\theta ,\phi ).
\end{equation*}

For the purpose of the main body article, we will use monopole harmonic
functions defined in momentum space.
To maintain the SU(2) algebra, the angular momentum operator
$\mathbf L$ in the presence of the monopole is augmented to
\bea
\mathbf L= \hbar \mathbf{k} \times \left(-i \mathbf \partial_k-\frac{1}{k} \mathbf{A}(\mathbf
k)\right)-q\hbar \hat{\mathbf{k}},
\eea
and the vector potential under the gauge chosen above is $\mathbf{A}(\mathbf
k)=q\tan\frac{\theta_k}{2}\hat e_{\phi_k}$.
We can simply replace polar and azimuthal angles of $\theta$ and $\phi$
of $\hat r$ with that of $\hat q$ in the expressions above to
yield ${\cal Y}_{q:jm}(\hat{\mathbf{k}})$.
Again for $\mathcal{Y}_{q;jm} (\hat{\mathbf{k}})$, $j$ starts with $|q|$ and takes values
of $|q|, |q|+1, ...$ For the lowest value $j=|q|$, $\mathcal{Y}%
_{q;j=|q|,m}(\hat{\mathbf{k}})$'s are holomorphic (anti-holomorphic) functions on the
unit sphere for $q<0$ ($q>0$).

\subsection{Ginzburg-Landau analysis}
\label{sect:GL}

To analyze the possible pairing configuration with $j=|q_p|=1$,
we perform the GL free energy analysis.
The GL free energy is constructed as
\begin{eqnarray}
F_{G}&=&\alpha |\Delta|^2 +\beta_1 |\Delta|^4 + \beta_2 |\Delta|^4
\sum_{mm^\prime}c_{1m}^* c_{1,-m}^* c_{1m^\prime} c_{1,-m^\prime},  \notag
\label{eq:GL}
\end{eqnarray}
in which $|\Delta|$ is the magnitude of gap and the normalized coefficients
$c_{1m}$ defined in Eq. (4) in the main text 
 control the angular distribution of the
gap function. Without loss of generality, $F_{GL}$ is minimized when
$c_{1m}$'s take the component of $m=1$ or $m=-1$ at $\beta_2>0$
and of $m=0$ at $\beta_2<0$, respectively.
Applying rotations to these configurations generates other equivalent
pairing configurations.
Following the standard terminology, we denote the former cases of
$m=\pm 1$ and the latter one of $m=0$ as the axial and the
polar pairing, respectively.

The Bogoliubov quasi-particle spectra become
\bea
E(\mathbf{k})=\sqrt{\epsilon_k^2 +\Delta^2(|k|) |f(u_k,v_k)|^2 }.
\eea
Assuming a fixed magnitude $\Delta(|k|)$ and normalized $f(\hat{\mathbf{k}})$, the
free energy functional depends on the location of nodes through the relation of
$F[u_k,v_k]=-\frac{2}{\beta}\sum_k \ln \left(2\cosh \beta E(\mathbf{k})\right)$.
Different from the vortex problem in real space, $F[u_k,v_k]$ does not
depend on the phase gradient in momentum space but only on the magnitude
distribution of $\Delta(\mathbf{k})$.
Because of the convexity of $F$, the two nodes repel each other
on $S_+$, and in the optimal configuration they lie at two ends
of a diameter, say, the north and south poles.
In other words, the mean-field results give rise to $\beta_2<0$, and
the corresponding pairing is polar with $m=0$.
Nevertheless, that case of $\beta_2>0$ cannot be ruled out.
The consequential axial pairing with $m=\pm 1$ will
a result beyond the mean-field theory arising from
strong correlation effect.

\subsection{Partial-wave analysis of interactions by monopole harmonics}
\label{sect:partialwave}

In this section, we elaborate the pair scattering interactions.
Since the inter-Fermi surface Cooper pairing is between two electrons
with parallel spins for the FS$_\pm$, we only consider the triplet
channel pairing whose Hamiltonian is expressed as
\bea
H_{pair}=\frac{1}{V_{0}}\sum_{\mathbf{k},\mathbf{k}^{\prime };m}V_{t}
(\mathbf{k},\mathbf{k}^{\prime })\chi _{1m}^{\dagger }(\mathbf{k})
\chi _{1m}(\mathbf{k}^{\prime })+h.c.,
\label{eq:int}
\eea
in which $V_{0}$ is the system volume; $\chi _{1m}^{\dagger }(\mathbf{k}%
)=\sum_{\sigma \sigma^{\prime }}\langle 1m|\frac{1}{2}\sigma;
\frac{1}{2}\sigma' \rangle c_{\sigma }^{\dagger }
(\mathbf{K}_{0}+\mathbf{k})c_{\sigma' }^{\dagger }(-\mathbf{K}_{0}-\mathbf{k})$
are the spin triplet pairing operator.
Because usual interactions in solids do not directly flip electron spins,
the spin index $m$ in Eq. (\ref{eq:int}) is expressed in the $S_{z}$%
-eigenbasis.
Eq. (\ref{eq:int}) is not expressed in the helical basis, and hence
it is still not the low energy pairing Hamiltonian in accommodation
to the helical Fermi surfaces yet.
After projecting the pairing Hamiltonian Eq. (\ref{eq:int}) into the
helical Fermi surface, we arrive at
\bea
{\tilde H}_{pair} = \sum_{\mathbf{k}, \mathbf{k}^\prime} {\tilde V}(\mathbf{k}, \mathbf
k^\prime) P^\dagger(\mathbf{k} ) P(\mathbf{k}^\prime)
+h.c.,
\eea
in which ${\tilde V}(\mathbf{k}, \mathbf{k}^\prime)
=\langle P(\mathbf{k}^{\prime })|H_{\text{pair}}|P(\mathbf{k})\rangle$
is the projected pair scattering matrix element.
It can be expressed as
\begin{eqnarray}
{\tilde V}(\mathbf{k}, \mathbf{k}^\prime) =\frac{4\pi}{3}\mathcal{Y}^*_{-1;1m}(\hat
{\mathbf{k}}) V_t(\mathbf{k}, \mathbf{k}^\prime) \mathcal{Y}_{-1;1,m}(\hat{\mathbf{k}}^\prime).
\label{eq:pj_int}
\end{eqnarray}
Here, $P(\mathbf{k}) =\sum_{m=-1}^{1}\sqrt{\frac{4\pi}{3}}
\mathcal{Y}_{-1;1m} (\hat{\mathbf{k}}) \chi_{1m} (\mathbf{k})$ was used
 and the rotational invariance of the interaction was assumed.
$\Delta(\mathbf{k})$ is the gap function determined by the
self-consistent equation
\bea
\Delta(\mathbf{k})=\frac{1}{V_0}\sum_{\mathbf{k}^\prime} {%
\tilde V} (\mathbf{k}, \mathbf{k}^\prime) \langle P(\mathbf{k}^\prime)\rangle.
\eea

Because of fermion statistics, the pair scattering matrix element in Eq. %
(\ref{eq:int}) is expressed as
\bea
V_{t}(\mathbf{k},\mathbf{k}^{\prime })=V(\mathbf{k}-%
\mathbf{k}^{\prime })-V(\mathbf{k}+\mathbf{k}^{\prime }+2\mathbf{K}_{0}),
\eea
in which the
first and second terms are the intra and inter-Fermi surface scattering,
respectively. If we neglect the inter-Fermi surface scattering which
involves large momentum transfer, $V_{t}(\mathbf{k},\mathbf{k}^{\prime })$ can be
assumed only depending on the relative angle between $\mathbf{k}$ and $\mathbf{k}%
^{\prime }$, i.e., $V_{t}(\mathbf{k},\mathbf{k}^{\prime })=V_{t}(\hat{\mathbf{k}}\cdot
\hat{\mathbf{k}}^{\prime })$. Unlike the usual case, i.e., $\mathbf{K}_{0}=0$, that $V_{t}$
only contains the odd partial-wave channels, here, both even and odd
partial-wave channels are allowed as $V(\hat{\mathbf{k}}\cdot \hat{\mathbf{k}}^{\prime
})=\sum_{lm}4\pi g_{l}
Y_{lm}^{\ast }(\hat{\mathbf{k}})Y_{lm}(\hat{\mathbf{k}})$ in which $l=0,1,2,...$.
As will be proved below, after the projection defined in Eq. (\ref%
{eq:pj_int}), the pairing interaction becomes
\begin{equation}
{\tilde{V}}(\hat{\mathbf{k}}\cdot \hat{\mathbf{k}}^{\prime })=\sum_{jm}\tilde{g}_{j}\mathcal{Y}%
_{-1,jm}^{\ast }(\hat{\mathbf{k}}^{\prime })\mathcal{Y}_{1,jm}(\hat{\mathbf{k}}),
\label{eq:intYqlm}
\end{equation}%
in which
$\tilde{g}_{j}=\frac{1}{2j+1}\sum_{l=j,j\pm 1}(2l+1)g_{l}|\langle
l0;11|j1\rangle |^{2}$,
and the partial wave channels start with $j=1$. In
other words, the projection to the helical Fermi surfaces reorganize the
partial-wave channels, and thus promoted the lowest partial wave channel
from $j=0$ to $j=1$. The actual pairing channel that the system takes is
determined by the most negative pairing matrix eigenvalue $V_{j^{\prime }}$.

Below, we prove Eq. (\ref{eq:intYqlm}). Starting from Eq. (\ref{eq:pj_int}),
since $V_t$ can be decomposed by usual spherical harmonics,
we have%
\begin{widetext}
\begin{equation}
{\tilde{V}}(\hat{\mathbf{k}}\cdot \hat{\mathbf{k}}^{\prime })=-\frac{(4\pi )^{2}}{3}%
\sum_{m_{1}}\mathcal{Y}_{-1;1m_{1}}^{\ast }(\hat{\mathbf{k}})\mathcal{Y}%
_{-1;1m_{1}}(\hat{\mathbf{k}}^{\prime
})[\sum_{l_{2}m_{2}}g_{l_{2}}Y_{l_{2}m_{2}}^{\ast }(\hat{\mathbf{k}}%
)Y_{l_{2}m_{2}}(\hat{\mathbf{k}}^{\prime })].
\label{eq:intmix}
\end{equation}%
where $g_{l_2}$ is the interaction strength of $V_t$ in the $l_2$-th partial-wave channel.
To further simplify ${\tilde{V}}(\hat{\mathbf{k}}\cdot \hat{\mathbf{k}}^{\prime })$, the
composition of monopole harmonics \cite{wu1976,wu1977} will be employed, which can
be derived by using D-matrices as follows: Let us start from the composition known for irreducible tensors of angular momentum,%
\begin{equation}
\Phi _{l_{1}l_{2}l_{3}m_{3}}=\sum_{m_{1}m_{2}}\Phi _{l_{1}m_{1}}\Phi
_{l_{2}m_{2}}\langle l_{3},m_{3}|l_{1},m_{1};l_{2},m_{2}\rangle ,
\label{eq:tensorcomposition1}
\end{equation}%
Alternatively,
\begin{equation}
\Phi _{l_{1}m_{1}}\Phi _{l_{2}m_{2}}=\sum_{l_{3}m_{3}}\Phi
_{l_{1}l_{2}l_{3}m_{3}}\langle l_{3},m_{3}|l_{1},m_{1};l_{2},m_{2}\rangle
\label{eq:tensorcomposition2}
\end{equation}%
Applying a rotation to both sides of Eq. (\ref{eq:tensorcomposition2}), we have
\begin{equation}
D(g)\Phi _{l_{1}m_{1}}\Phi _{l_{2}m_{2}}=\sum_{l_{3}m_{3}}D(g)\Phi
_{l_{1}l_{2}l_{3}m_{3}}\langle l_{3},m_{3}|l_{1},m_{1};l_{2},m_{2}\rangle .
\end{equation}%
It can be represented by D-matrices as%
\begin{equation}
\sum_{m_{1}^{\prime }m_{2}^{\prime }}\Phi _{l_{1}m_{1}^{\prime }}\Phi
_{l_{2}m_{2}^{\prime }}D_{m_{1}^{\prime },m_{1}}^{l_{1}}(g)D_{m_{2}^{\prime
},m_{2}}^{l_{2}}(g)=\sum_{l_{3}m_{3}m_{3}^{\prime }}\Phi
_{l_{1}l_{2}l_{3}m_{3}^{\prime }}D_{m_{3}^{\prime },m_{3}}^{l_{3}}(g)\langle
l_{3},m_{3}|l_{1},m_{1};l_{2},m_{2}\rangle ,
\end{equation}%
Using Eq. (\ref{eq:tensorcomposition1}), the right hand side of the above equation becomes
\begin{equation}
\sum_{l_{3}m_{3}m_{3}^{\prime }}\sum_{m_{1}^{\prime
}m_{2}^{\prime }}\Phi _{l_{1}m_{1}^{\prime }}\Phi _{l_{2}m_{2}^{\prime
}}D_{m_{3}^{\prime },m_{3}}^{l_{3}}(g)\langle l_{3},m_{3}^{\prime
}|l_{1},m_{1}^{\prime };l_{2},m_{2}^{\prime }\rangle \langle
l_{3},m_{3}|l_{1},m_{1};l_{2},m_{2}\rangle .
\end{equation}%
Therefore, the composition rule of D-matrices is obtained as%
\begin{equation}
D_{m_{1}^{\prime },m_{1}}^{l_{1}}(g)D_{m_{2}^{\prime
},m_{2}}^{l_{2}}(g)=\sum_{l_{3}m_{3}m_{3}^{\prime }}\langle
l_{3},m_{3}^{\prime }|l_{1},m_{1}^{\prime };l_{2},m_{2}^{\prime }\rangle
\langle l_{3},m_{3}|l_{1},m_{1};l_{2},m_{2}\rangle D_{m_{3}^{\prime
},m_{3}}^{l_{3}}(g).
\end{equation}%
By changing variables, we have%
\begin{equation}
D_{m_{1},-q_{1}}^{l_{1}\ast }(g)D_{m_{2},-q_{2}}^{l_{2}\ast
}(g)=\sum_{l_{3}=|l_{1}-l_{2}|}^{l_{1}+l_{2}}\langle
l_{3},m_{3}|l_{1},m_{1};l_{2},m_{2}\rangle \langle
l_{3},-q_{3}|l_{1},-q_{1};l_{2},-q_{2}\rangle D_{m_{3},-q_{3}}^{l_{3}\ast
}(g),
\end{equation}%
where $m_{3}=m_{1}+m_{2}$, $q_{3}=q_{1}+q_{2}$. Using Eq. (\ref{eq:monoDmatrix}), the composition of monopole harmonics can be derived from the above composition of D-matrices as%
\begin{eqnarray}
\sqrt{\frac{4\pi }{2l_{1}+1}}\sqrt{\frac{4\pi }{2l_{2}+1}}\mathcal{Y}%
_{q_{1};l_{1}m_{1}}(\hat{\mathbf{k}})\mathcal{Y}_{q_{2};l_{2}m_{2}}(\hat{\mathbf{k}})
&=& \sum_{l_{3}}\langle l_{3},m_{3}|l_{1},m_{1};l_{2},m_{2}\rangle \langle
l_{3},-q_{3}|l_{1},-q_{1};l_{2},-q_{2}\rangle \sqrt{\frac{4\pi }{2l_{3}+1}}%
\mathcal{Y}_{q_{3};l_{3}m_{3}}(\hat{\mathbf{k}}). \nn \\
\end{eqnarray}%
For our derivation, let us take a special case of $q_{1}=-1$, $q_{2}=0$, and
$l_{1}=1$,
\begin{equation}
\mathcal{Y}_{-1;1m_{1}}(\hat{\mathbf{k}})Y_{l_{2}m_{2}}(\hat{\mathbf{k}})=%
\sum_{l_{3}=l_{2}-1}^{l_{2}+1}\sqrt{\frac{3(2l_{2}+1)}{4\pi (2l_{3}+1)}}%
\langle l_{3},m_{3}|1,m_{1};l_{2},m_{2}\rangle \langle
l_{3},1|1,1;l_{2},0\rangle \mathcal{Y}_{-1;l_{3}m_{3}}(\hat{\mathbf{k}}).
\end{equation}%
Then, Eq. (\ref{eq:intmix}) can be simplified as%
\begin{eqnarray}
{\tilde{V}}(\hat{\mathbf{k}}\cdot \hat{\mathbf{k}}^{\prime }) &=&-4\pi
\sum_{l_{2}}g_{l_{2}}\sum_{m_{1},m_{2}}%
\sum_{l_{3}=l_{2}-1}^{l_{2}+1}\sum_{l_{3}^{\prime }=l_{2}-1}^{l_{2}+1}\frac{%
2l_{2}+1}{\sqrt{(2l_{3}+1)(2l_{3}^{\prime }+1)}} \times \nn \\
&&\times \langle l_{3},m_{3}|1,m_{1};l_{2},m_{2}\rangle \langle
l_{3},1|1,1;l_{2},0\rangle \mathcal{Y}_{-1;l_{3}m_{3}}^{\ast }(\hat{\mathbf{k}}) \times \nn \\
&&\times \langle l_{3}^{\prime },m_{3}|1,m_{1};l_{2},m_{2}\rangle \langle
l_{3}^{\prime },1|1,1;l_{2},0\rangle \mathcal{Y}_{-1;l_{3}^{\prime }m_{3}}(%
\hat{\mathbf{k}}^{\prime }).
\end{eqnarray}%
The orthogonality of Clebsch-Gordan coefficients $\sum_{m_{2}}\langle
l_{3},m_{3}|1,m_{3}-m_{2};l_{2},m_{2}\rangle \langle l_{3}^{\prime
},m_{3}|1,m_{3}-m_{2};l_{2},m_{2}\rangle =\delta _{l_{3}l_{3}^{\prime }}$
further gives
\begin{eqnarray}
{\tilde{V}}(\hat{\mathbf{k}}\cdot \hat{\mathbf{k}}^{\prime }) &=&-4\pi
\sum_{l_{2}}g_{l_{2}}\sum_{l_{3}=l_{2}-1}^{l_{2}+1}\frac{2l_{2}+1}{%
2l_{3}+1}\langle l_{3},1|1,1;l_{2},0\rangle ^{2}\sum_{m_{3}}\mathcal{%
Y}_{-1;l_{3}m_{3}}^{\ast }(\hat{\mathbf{k}})\mathcal{Y}_{-1;l_{3}m_{3}}(\hat{\mathbf{k}}%
^{\prime }) \nn \\
&=&-4\pi \sum_{l_{3}m_{3}}\tilde{g}_{l_{3}}\mathcal{Y}%
_{-1;l_{3}m_{3}}^{\ast }(\hat{\mathbf{k}})\mathcal{Y}_{-1;l_{3}m_{3}}(\hat{\mathbf{k}}^{\prime
}),
\end{eqnarray}
\end{widetext}
where $\tilde{g}_{l_{3}}=\frac{1}{2l_{3}+1}%
\sum_{l_{2}=l_{3}-1}^{l_{3}+1}(2l_{2}+1)g_{l_{2}}\langle
l_{3},1|1,1;l_{2},0\rangle ^{2}$ as shown in Eq. (\ref{eq:intYqlm}).

\subsection{Intra-Fermi surface pairing}
\label{sect:FFLO}

Another pairing possibility is the intra Fermi surface pairing:
Cooper pairings take place within each Fermi surface FS$_\pm$
carrying finite momentum $\pm \mathbf{K}_0$,
respectively, thus this is an example of the Fulde-Ferrel-Larkin-Ovchinnikov
type pairing \cite{larkin1964,fulde1964}.
The intra FS$_\pm$ pairing operators are defined as
\bea
P^\dagger_+(\mathbf{k})=\alpha^\dagger_+(\mathbf{k}) \alpha^\dagger_+(-\mathbf{k}),
P^\dagger_-(\mathbf{k})=\alpha^\dagger_-(\mathbf{k}) \alpha^\dagger_-(-\mathbf{k}),
\eea
which satisfy $[\mathbf{S} \cdot \hat{\mathbf{k}}, P^\dagger_\pm(\mathbf{k})]=0$.
Therefore, their pairing phases can be well-defined globally over
the sphere $|k|=k_f$ without non-trivial monopole structure,
such that the gap functions can be decomposed into the usual
spherical harmonic functions as $\Delta_\pm (\mathbf
k) =\Delta_{jm} (|k|)\sum_{jm} c_{\pm, jm} \mathcal{Y}_{jm}(\hat{\mathbf{k}})$.
In the simplest case with $j=m=0$, the pairing $\Delta_{0,\pm}$ is a mixture
between the conventional $s$-wave and $^3$He-B type $p$-wave pairings.
Under the $\sigma_z$-eigen basis, they are written as
\bea
P_{\pm}(\mathbf{k})\propto
\chi^\dagger_{\pm;00}(\mathbf{k}) \pm \mathbf{k} \cdot \mathbf \chi_{\pm}(\mathbf{k}),
\eea
in which the singlet and triplet pairing operators are
$\chi^\dagger_{\pm;00}(\mathbf{k})=\sum_{\alpha\beta} \langle 00|\frac{1}{2%
}\alpha\frac{1}{2}\beta\rangle c^\dagger_\alpha(\pm \mathbf{K}_0+\mathbf{k})
c^\dagger_\beta(\pm \mathbf{K}_0-\mathbf{k})$, and $\mathbf \chi_{\pm}(\mathbf{k})=
c^\dagger_\alpha(\pm \mathbf{K}_0+\mathbf{k}) i(\sigma_y \mathbf \sigma)_{\alpha\beta}
c^\dagger_\beta(\pm \mathbf{K}_0-\mathbf{k})$.
This pairing is fully gapped, and thus can be characterized by the
integer-valued topological index in the DIII class
for 3D superconductivity.
According to the criterion in Ref. [\onlinecite{qi2010}],
the overall pairing structure is topologically non-trivial according
to $\Delta_+=\mp \Delta_-$, respectively.
The competition between the inter- and intra-Fermi surface pairing is an
interesting question depending on the concrete pairing mechanism.

\end{document}